\documentclass{appolb}
\usepackage{cite}
\usepackage[centertags]{amsmath}
\allowdisplaybreaks[4]
\numberwithin{equation}{section}
\usepackage{amsbsy}
\usepackage{amsfonts}
\usepackage{amssymb}
\usepackage[dvips]{graphicx}
\usepackage{url}
\usepackage[dvips]{hyperref}
\begin{document}


\title{Absolute Neutrino Masses\thanks{
Talk presented at the
XXIX International Conference of Theoretical Physics
``Matter To The Deepest: Recent Developments In Physics Of Fundamental Interactions'',
8-14 September 2005, Ustron, Poland.
}
}
\author{Carlo Giunti
\address{INFN, Sezione di Torino, and Dipartimento di Fisica Teorica,
Universit\`a di Torino,
Via P. Giuria 1, I--10125 Torino, Italy}
}
\maketitle
\begin{abstract}
The main aspects of the phenomenology of absolute neutrino masses are reviewed,
focusing on the limits on neutrino masses obtained in
tritium $\beta$ decay experiments,
cosmological observations
and
neutrinoless double-$\beta$ decay experiments.
\end{abstract}

\section{Introduction}
\label{Introduction}

Neutrino oscillation experiments
have shown that neutrinos are massive and mixed particles,
\textit{i.e.}
the left-handed components
$\nu_{\alpha L}$
of flavor neutrinos
($\alpha=e,\mu,\tau$)
are linear combinations of
the left-handed components
$\nu_{k L}$
of neutrinos with masses $m_k$:
\begin{equation}
\nu_{\alpha L} = \sum_{k} U_{\alpha k} \, \nu_{kL}
\,,
\label{001}
\end{equation}
where $U$ is the mixing matrix
(see Refs.~\cite{hep-ph/9812360,hep-ph/0211462,hep-ph/0310238}).
Neutrino oscillations depend on
the elements of the mixing matrix $U$,
which determine the amplitude of the oscillations,
and on the squared-mass differences
$ \Delta{m}^2_{kj} \equiv m_k^2 - m_j^2 $,
which determine the oscillation length.

The results of
solar neutrino experiments
(Homestake \cite{Cleveland:1998nv},
Kamiokande \cite{Fukuda:1996sz},
SAGE \cite{astro-ph/0204245},
GALLEX \cite{Hampel:1998xg},
GNO \cite{hep-ex/0504037},
Super-Kamiokande \cite{hep-ex/0508053}
and
SNO \cite{nucl-ex/0502021})
and the reactor long-baseline experiment KamLAND \cite{hep-ex/0406035}
imply that there are large
$ \nu_e \to \nu_{\mu}, \nu_{\tau} $
transitions with a squared-mass difference
$ \Delta{m}^2_{\mathrm{SUN}} \simeq 8 \times 10^{-5} \, \text{eV}^2 $.

Atmospheric neutrino experiments
(Kamiokande \cite{Fukuda:1994mc},
IMB \cite{Becker-Szendy:1992hq},
Super-Kamiokande \cite{hep-ex/0501064},
Soudan-2 \cite{hep-ex/0507068}
and
MACRO \cite{hep-ex/0304037})
and the accelerator K2K experiment \cite{hep-ex/0411038},
together with the negative results of the CHOOZ experiment \cite{hep-ex/0301017},
have shown that there are large
$ \nu_{\mu} \to \nu_{\tau} $
transitions with a squared-mass difference
$ \Delta{m}^2_{\mathrm{ATM}} \simeq 2.5 \times 10^{-3} \, \text{eV}^2 $.

\begin{figure}[t!]
\begin{center}
\begin{minipage}[t]{0.6\textwidth}
\begin{tabular}{lr}
\includegraphics*[bb=183 469 425 771, width=0.45\linewidth]{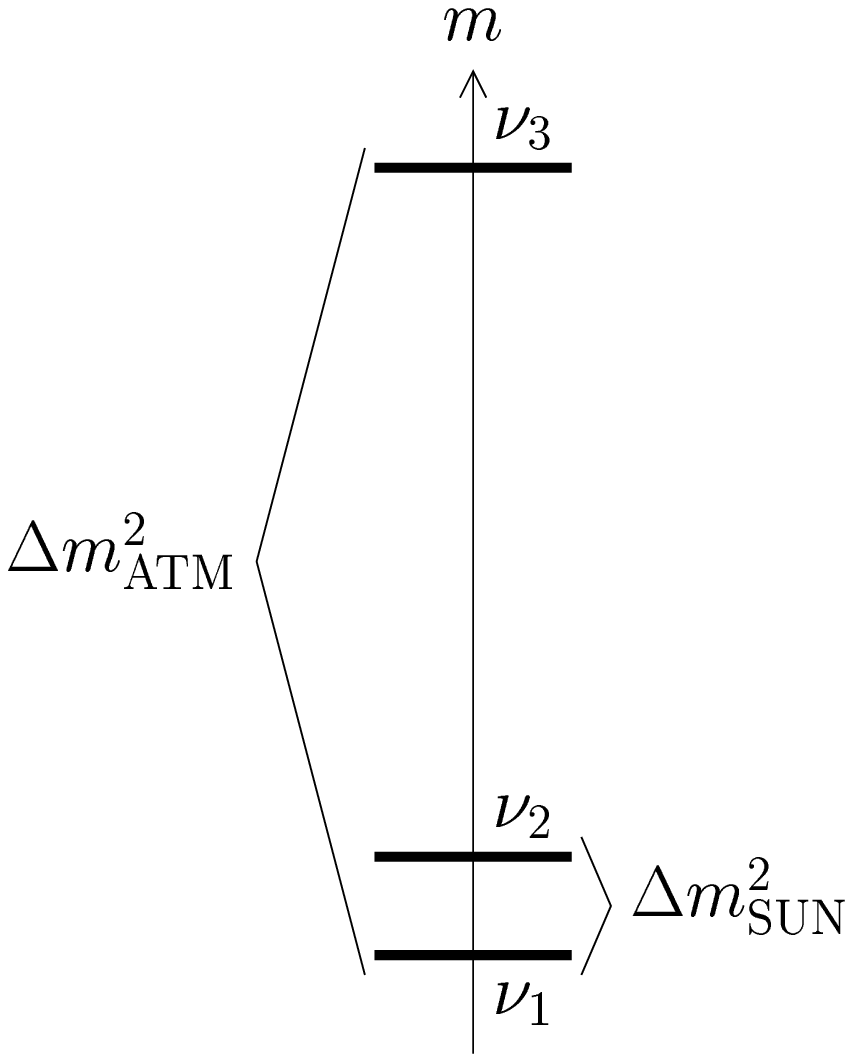}
&
\includegraphics*[bb=186 469 428 771, width=0.45\linewidth]{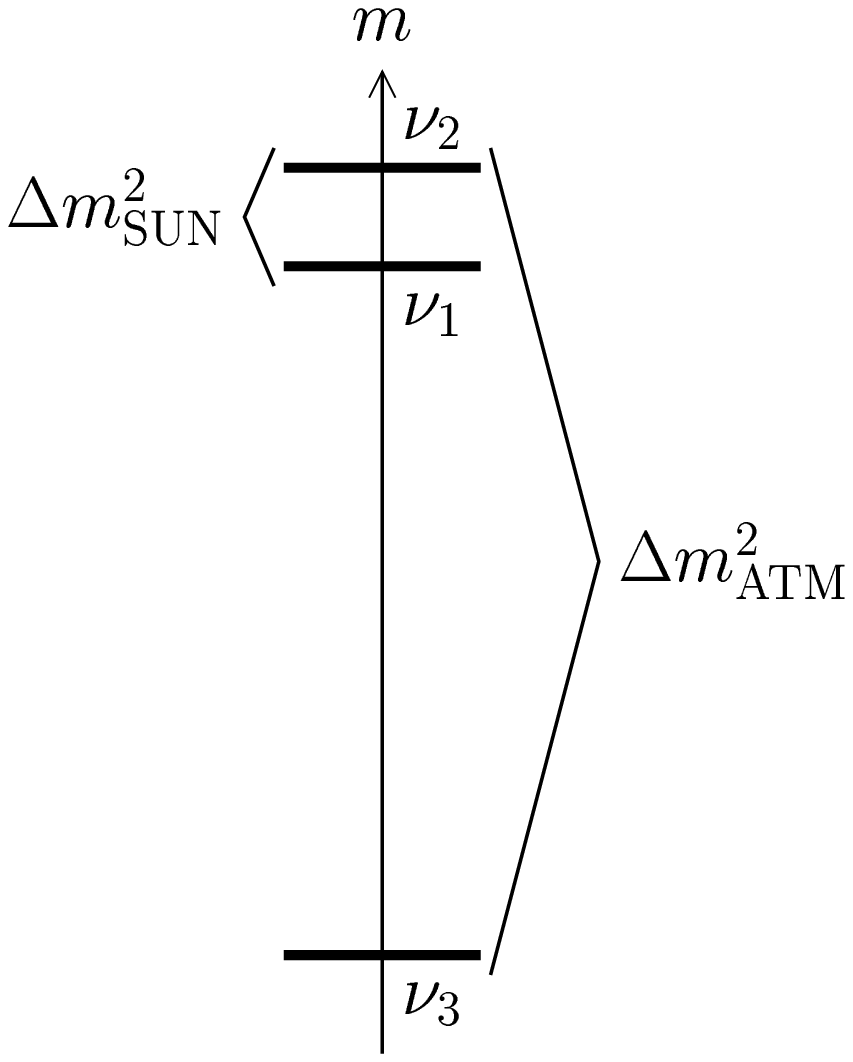}
\\
\textbf{normal}
&
\textbf{inverted}
\end{tabular}
\end{minipage}
\end{center}
\caption{ \label{3nu}
The two three-neutrino schemes allowed by the hierarchy
$\Delta{m}^2_{\text{SUN}} \ll \Delta{m}^2_{\text{ATM}}$.
}
\end{figure}

Since
$ \Delta{m}^2_{\mathrm{ATM}} \gg \Delta{m}^2_{\mathrm{SUN}} $,
at least two independent squared-mass differences are needed in order to explain
the results of neutrino oscillation experiments.
This requirement is satisfied in the simplest case of
three-neutrino mixing, in which the number of massive neutrinos
in Eq.~(\ref{001}) is three
(see Refs.~\cite{hep-ph/9812360,hep-ph/0211462,hep-ph/0310238}).
In such a framework,
there are two types of possible schemes,
which are shown in Fig.~\ref{3nu}.
We labeled the massive neutrinos in order to have
$\Delta{m}^2_{21}=\Delta{m}^2_{\mathrm{SUN}}$
and
$|\Delta{m}^2_{31}|=\Delta{m}^2_{\mathrm{ATM}}$,
with $ \Delta{m}^2_{32} \simeq \Delta{m}^2_{31} $.
In the normal scheme, which is so-called because it allows a mass hierarchy
$m_1 \ll m_2 \ll m_3$,
the squared-mass difference
$\Delta{m}^2_{31}$ is positive,
whereas in the inverted scheme
it is negative.

A global fit of the oscillation data \cite{hep-ph/0408045} gives the best-fits
and
$3\sigma$
ranges for the three-neutrino oscillation parameters listed in Tab.~\ref{fit}.
The mixing angles
$\vartheta_{12}$,
$\vartheta_{13}$,
$\vartheta_{23}$
belong to the standard parameterization of the mixing matrix
\cite{Eidelman:2004wy},
in which, with good approximation,
$\vartheta_{12}$
is the solar mixing angle,
$\vartheta_{23}$
is the atmospheric mixing angle,
and
$\vartheta_{13}$
is the CHOOZ mixing angle
\cite{Bilenky:1998tw,hep-ph/0212142}.
In Tab.~\ref{fit} we give only the values of
$\vartheta_{12}$
and
$\vartheta_{13}$,
which are sufficient for the following discussion
on the phenomenology of absolute neutrino masses.

\begin{table}[b!]
\caption{ \label{fit} Best-fit
and
$3\sigma$
range for the three-neutrino oscillation parameters
obtained in the global fit of Ref.~\cite{hep-ph/0408045}.
}
\begin{center}
\begin{tabular}{|c|c|}
\hline
Parameter & \begin{tabular}{c} Best-Fit \\ $3\sigma$ Range \end{tabular}
\\
\hline
$\Delta{m}^2_{21}$
&
\begin{tabular}{c} 
$8.3 \times 10^{-5} \, \mathrm{eV}^2$
\\
$7.4 \times 10^{-5} \, - \, 9.3 \times 10^{-5} \, \mathrm{eV}^2$
\end{tabular}
\\
\hline
$\sin^2\vartheta_{12}$
&
\begin{tabular}{c} 
$0.28$
\\
$0.22 \, - \, 0.37$
\end{tabular}
\\
\hline
$|\Delta{m}^2_{31}|$
&
\begin{tabular}{c} 
$2.4 \times 10^{-3} \, \mathrm{eV}^2$
\\
$1.8 \times 10^{-3} \, - \, 3.2 \times 10^{-3} \, \mathrm{eV}^2$
\end{tabular}
\\
\hline
$\sin^2\vartheta_{13}$
&
\begin{tabular}{c} 
$0.01$
\\
$0 \, - \, 0.05$
\end{tabular}
\\
\hline
\end{tabular}
\end{center}
\end{table}

\begin{figure}[t!]
\begin{minipage}[t]{0.47\textwidth}
\begin{center}
\includegraphics*[bb=120 427 463 750, width=0.95\textwidth]{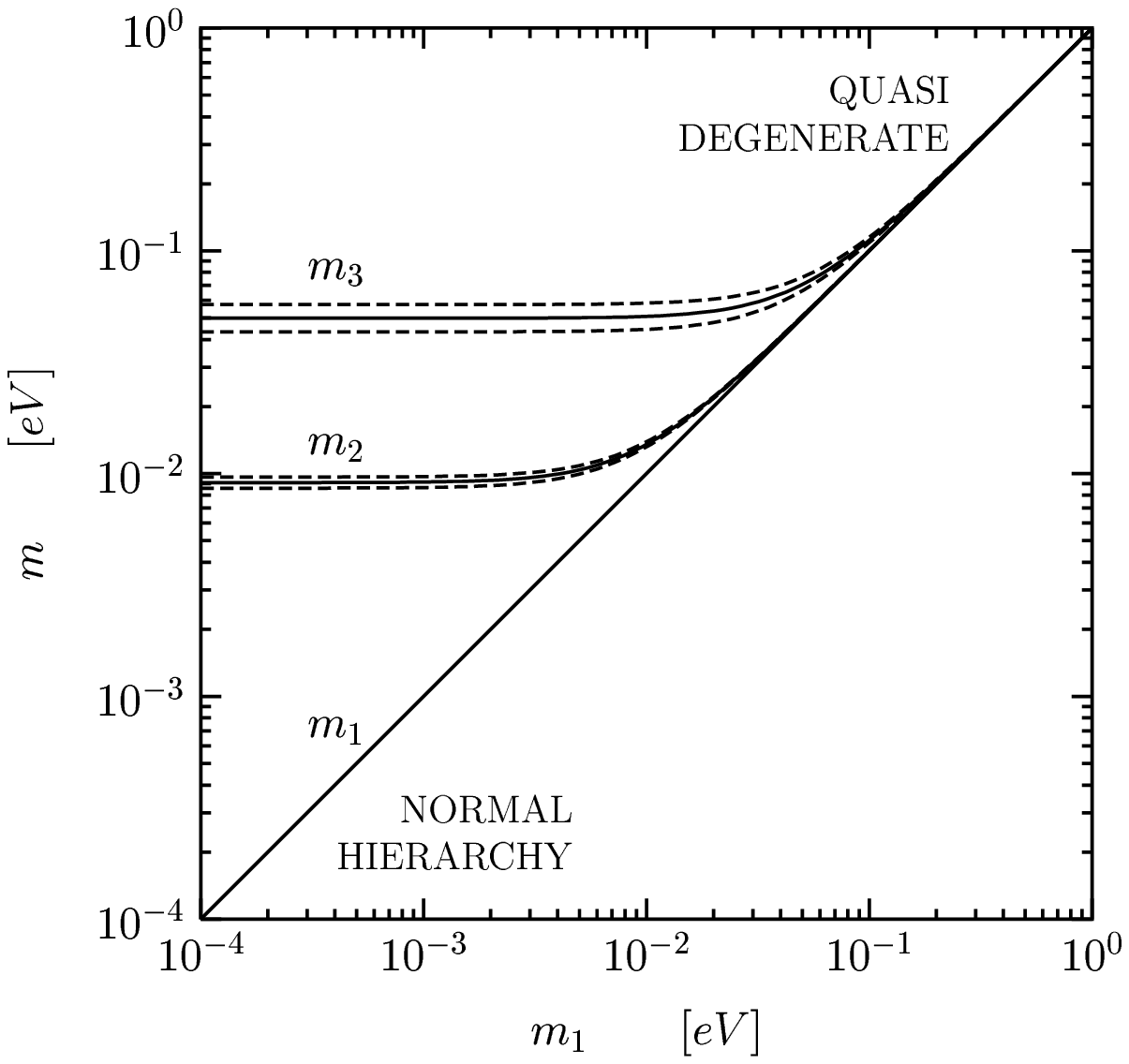}
\end{center}
\end{minipage}
\hfill
\begin{minipage}[t]{0.47\textwidth}
\begin{center}
\includegraphics*[bb=120 427 463 750, width=0.95\textwidth]{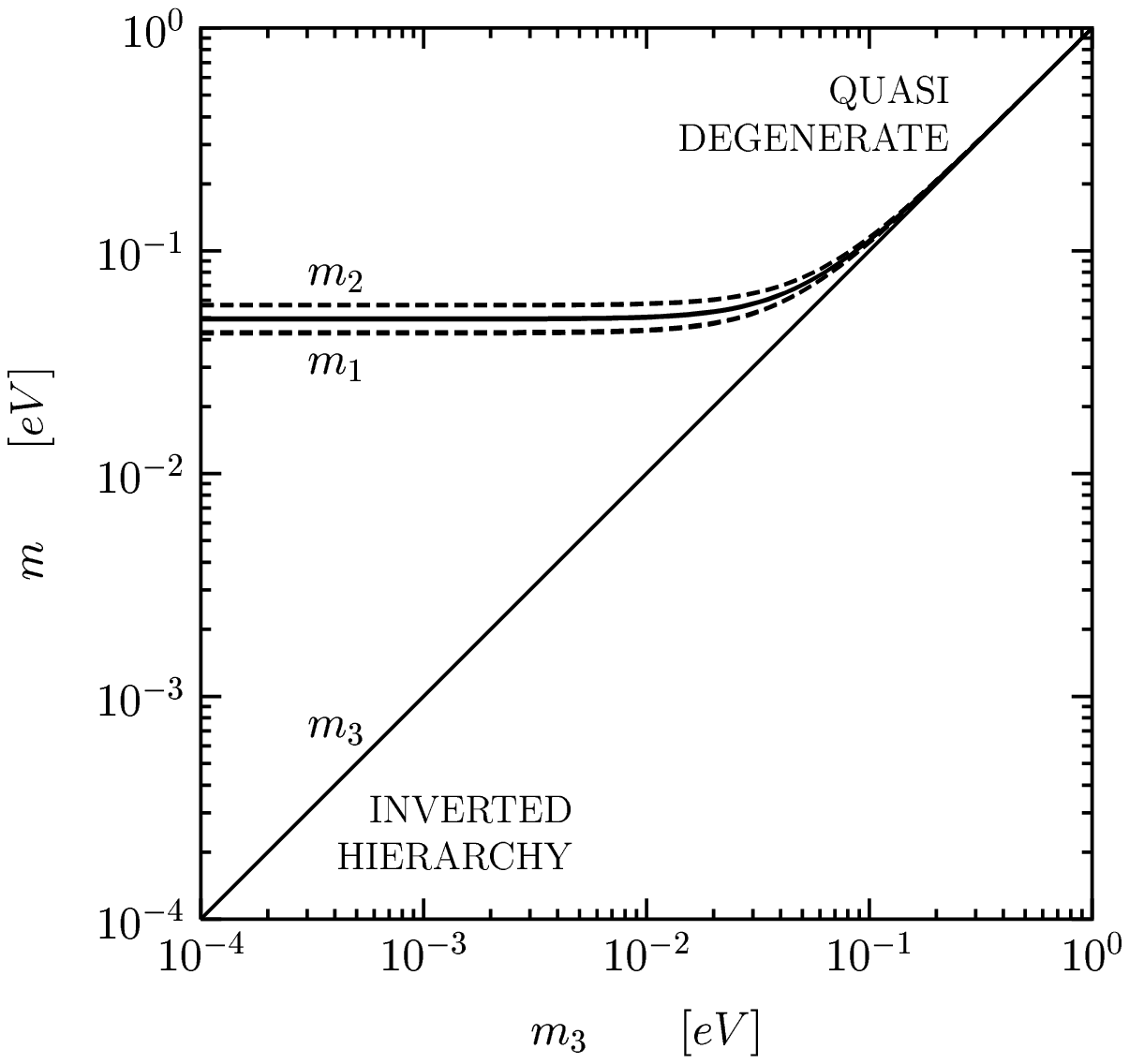}
\end{center}
\end{minipage}
\caption{ \label{3ma}
Values of neutrino masses as functions
of the lightest mass
$m_1$ in the normal scheme and $m_3$ in the inverted scheme.
Solid lines correspond to the best-fit in Tab.~\ref{fit}.
Dashed lines enclose $3\sigma$ ranges.
}
\end{figure}

Neutrino oscillations depend on the differences of neutrino masses,
not on their absolute values.
As we will see in the following,
other experiments
are able to give information on the absolute values of neutrino masses.
Figure~\ref{3ma} shows the values of
the neutrino masses
obtained from
$\Delta{m}^2_{21}$ and $|\Delta{m}^2_{31}|$ in Tab.~\ref{fit}
as functions
of the unknown value of the lightest mass,
which is $m_1$ in the normal scheme
and
$m_3$ in the inverted scheme.
As shown in the figure,
the case $m_3 \ll m_1 \lesssim m_2$
is conventionally called ``inverted hierarchy'',
whereas
in both normal and inverted schemes we have quasi-degeneracy of
neutrino masses for
$
m_1 \simeq m_2 \simeq m_3
\gg
\sqrt{\Delta{m}^2_{\mathrm{ATM}}} \simeq 5 \times 10^{-2} \, \mathrm{eV}
$.
In the inverted scheme
$\nu_1$ and $\nu_2$
are quasi-degenerate for any value of $m_{3}$
and their best-fits values and $3\sigma$ ranges
are practically superimposed in Figure~\ref{3ma}.

In the following, we review the phenomenology
of absolute neutrino masses in
tritium $\beta$ decay (Section~\ref{Tritium}),
cosmological measurements (Section~\ref{Cosmological})
and
neutrinoless double-$\beta$ decay (Section~\ref{Neutrinoless}).

\section{Tritium $\beta$ Decay}
\label{Tritium}

The measurement of the electron spectrum
in $\beta$ decays provides a robust direct determination
of the values of neutrino masses.
In practice, the most sensitive experiments
use tritium $\beta$ decay,
because it is a super-allowed transition with a low $Q$-value.
Information on neutrino masses is obtained by measuring the
Kurie function $K(T)$, given by
\cite{Shrock:1980vy,McKellar:1980cn,Kobzarev:1980nk}
\begin{equation}
K^2(T)
=
( Q - T )
\sum_k
|U_{ek}|^2
\sqrt{ (Q-T)^2 - m_k^2 }
\,,
\label{002}
\end{equation}
where $T$ is the electron kinetic energy.
The effect of neutrino masses
can be observed
near the end point of the electron spectrum,
where
$ Q-T \sim m_k $.
A low $Q$-value is important, because
the relative number of events occurring in an interval of energy
$\Delta{T}$
near the end-point is proportional to $(\Delta{T}/Q)^3$.

In the case of three-neutrino mixing,
the Kurie function in Eq.~(\ref{002})
depends on three neutrino masses and two mixing parameters
(the unitarity of the mixing matrix implies that
$ \sum_k |U_{ek}|^2 = 1 $).
However,
since so far tritium $\beta$ decay experiments did not see
any effect due to neutrino masses,
it is possible to approximate
$m_k \ll Q-T$
and obtain
\begin{equation}
K^2(T)
\simeq
( Q - T )
\sqrt{ (Q-T)^2 - m_{\beta}^2 }
\,.
\label{g029}
\end{equation}
This is a function of only one parameter,
the effective neutrino mass $m_{\beta}$, given by
\cite{Shrock:1980vy,McKellar:1980cn,Kobzarev:1980nk,Holzschuh:1992xy,Weinheimer:1999tn,Vissani:2000ci,hep-ph/0211341}
\begin{equation}
m_{\beta}^2 = \sum_k |U_{ek}|^2 m_k^2
\,.
\label{g030}
\end{equation}

The current best upper bounds on $m_{\beta}$ have been obtained in the
Mainz and Troitsk experiments
(see Ref.~\cite{hep-ex/0210050}):
\begin{equation}
m_\beta < 2.2 \, \mathrm{eV}
\qquad
\text{(95\% CL)}
\,.
\label{g032}
\end{equation}
In the near future, the KATRIN experiment \cite{hep-ex/0309007}
will reach a sensitivity of about $0.2 \, \mathrm{eV}$.

\begin{figure}[t!]
\begin{minipage}[t]{0.47\textwidth}
\begin{center}
\includegraphics*[bb=102 430 445 754, width=0.95\textwidth]{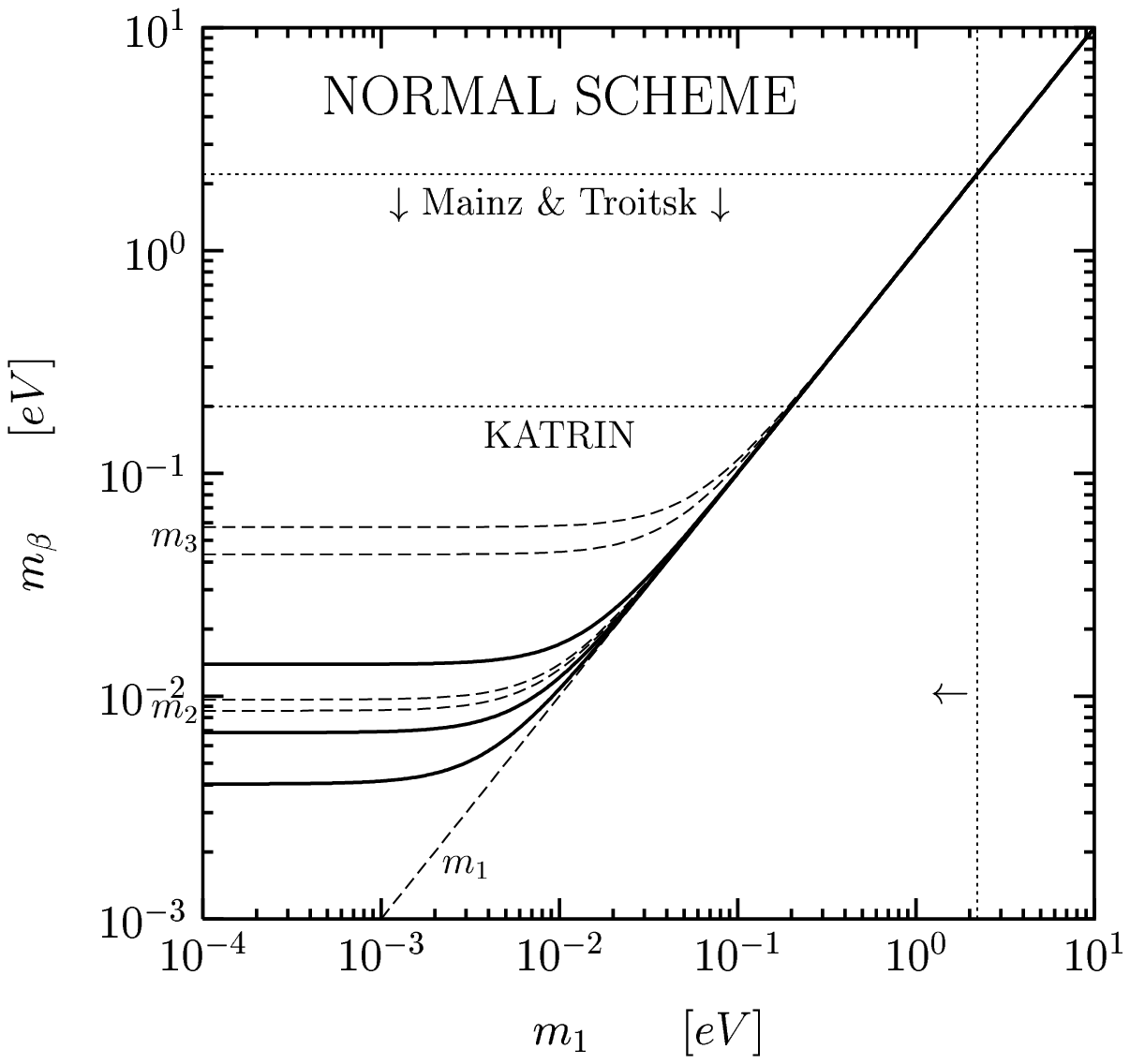}
\end{center}
\end{minipage}
\hfill
\begin{minipage}[t]{0.47\textwidth}
\begin{center}
\includegraphics*[bb=102 430 445 754, width=0.95\textwidth]{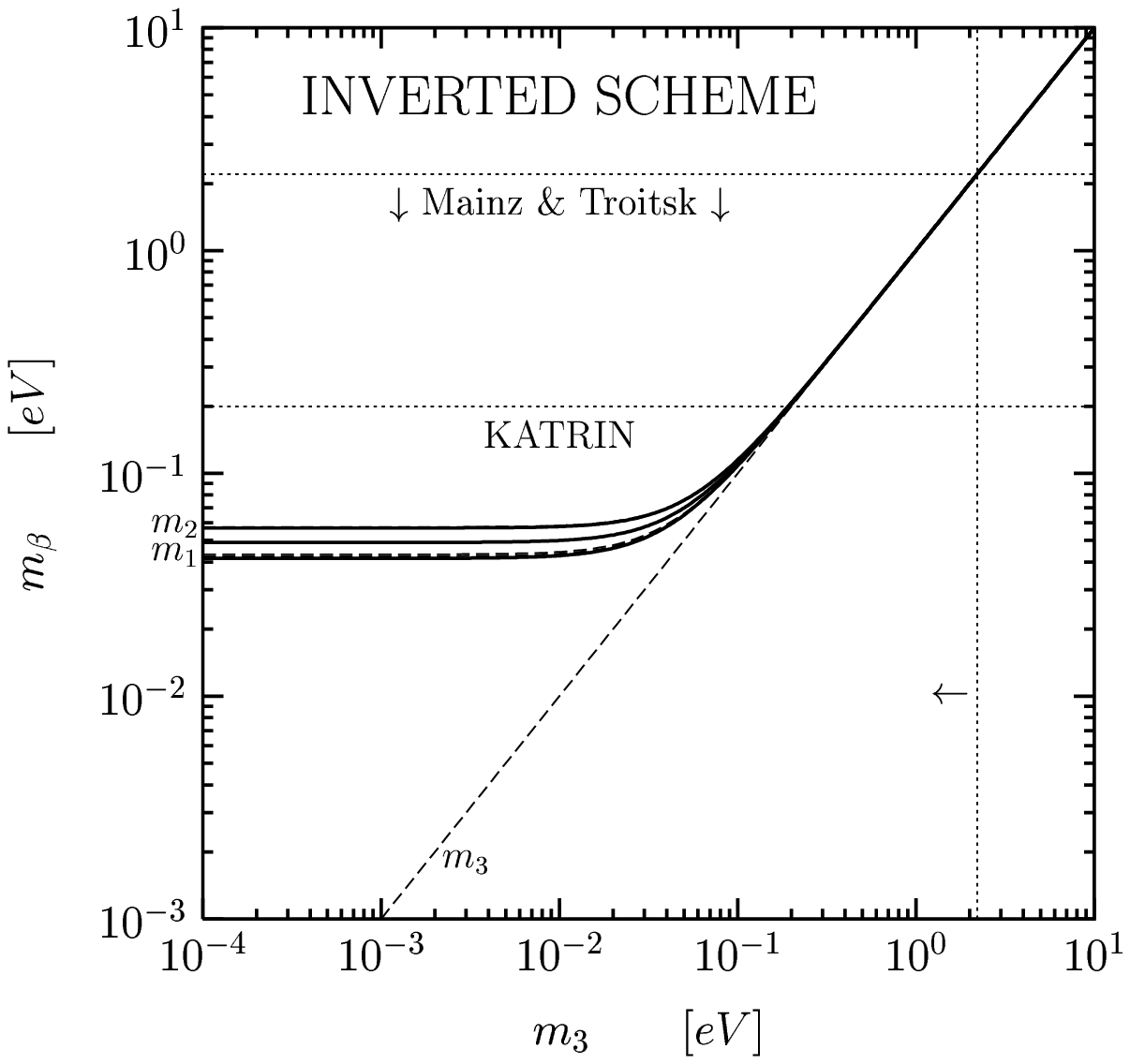}
\end{center}
\end{minipage}
\caption{ \label{mb}
Effective neutrino mass $m_\beta$
in tritium $\beta$-decay experiments as a function
of the lightest mass $m_1$ in the normal scheme and $m_3$ in the inverted scheme.
Middle solid lines correspond to the best-fit in Tab.~\ref{fit}.
Extreme solid lines enclose $3\sigma$ ranges.
Dashed lines delimit $3\sigma$ ranges of individual masses.
}
\end{figure}

In the standard parameterization of the mixing matrix we have
($c_{ij} \equiv \cos\vartheta_{ij}$
and
$s_{ij} \equiv \sin\vartheta_{ij}$)
\begin{equation}
m_{\beta}^2
=
c_{12}^2 \, c_{13}^2 \, m_1^2
+
s_{12}^2 \, c_{13}^2 \, m_2^2
+
s_{13}^2 \, m_3^2
\,.
\label{g031}
\end{equation}
Since the values of
$\Delta{m}^2_{21}$,
$|\Delta{m}^2_{31}|$,
$\vartheta_{12}$
and
$\vartheta_{13}$
are determined by neutrino oscillation experiments,
there is only one unknown quantity in Eq.(\ref{g031}),
which corresponds to the absolute scale of neutrino masses.
Figure~\ref{mb} shows the value of $m_{\beta}$
as a function
of the unknown value of the lightest mass
($m_1$ in the normal scheme
and
$m_3$ in the inverted scheme),
using the values of the oscillation parameters
in Tab.~\ref{fit}.
The middle solid lines correspond to the best fit
and the extreme solid lines delimit the $3\sigma$ allowed range.
We have also shown with dashed lines
the $3\sigma$ ranges of the neutrino masses
(same as in Fig.~\ref{3ma}),
which help to understand their contribution to
$m_{\beta}$.
One can see that, in the case of a normal mass hierarchy
(normal scheme with
$m_1 \ll m_2 \ll m_3$),
the main contribution to $m_{\beta}$
is due to $m_2$ or $m_3$ or both,
because the upper limit for
$m_{\beta}$
is larger than the upper limit for
$m_2$.
In the case of an inverted mass hierarchy
(inverted scheme with
$m_3 \ll m_1 \lesssim m_2$),
$m_{\beta}$
has practically the same value as $m_1$ and $m_2$.

Figure~\ref{mb}
shows that the Mainz and Troitsk experiments
and the near-future KATRIN experiment
give information on the absolute values of neutrino masses
in the quasi-degenerate region
in both normal and inverted schemes.
In the far future, the inverted scheme could be excluded
if experiments with a sensitivity of
about
$4 \times 10^{-2} \, \mathrm{eV}$
will not find any effect of neutrino masses.

\section{Cosmological Measurements}
\label{Cosmological}

\begin{figure}[t!]
\begin{minipage}[t]{0.47\textwidth}
\begin{center}
\includegraphics*[bb=102 430 437 748, width=0.95\textwidth]{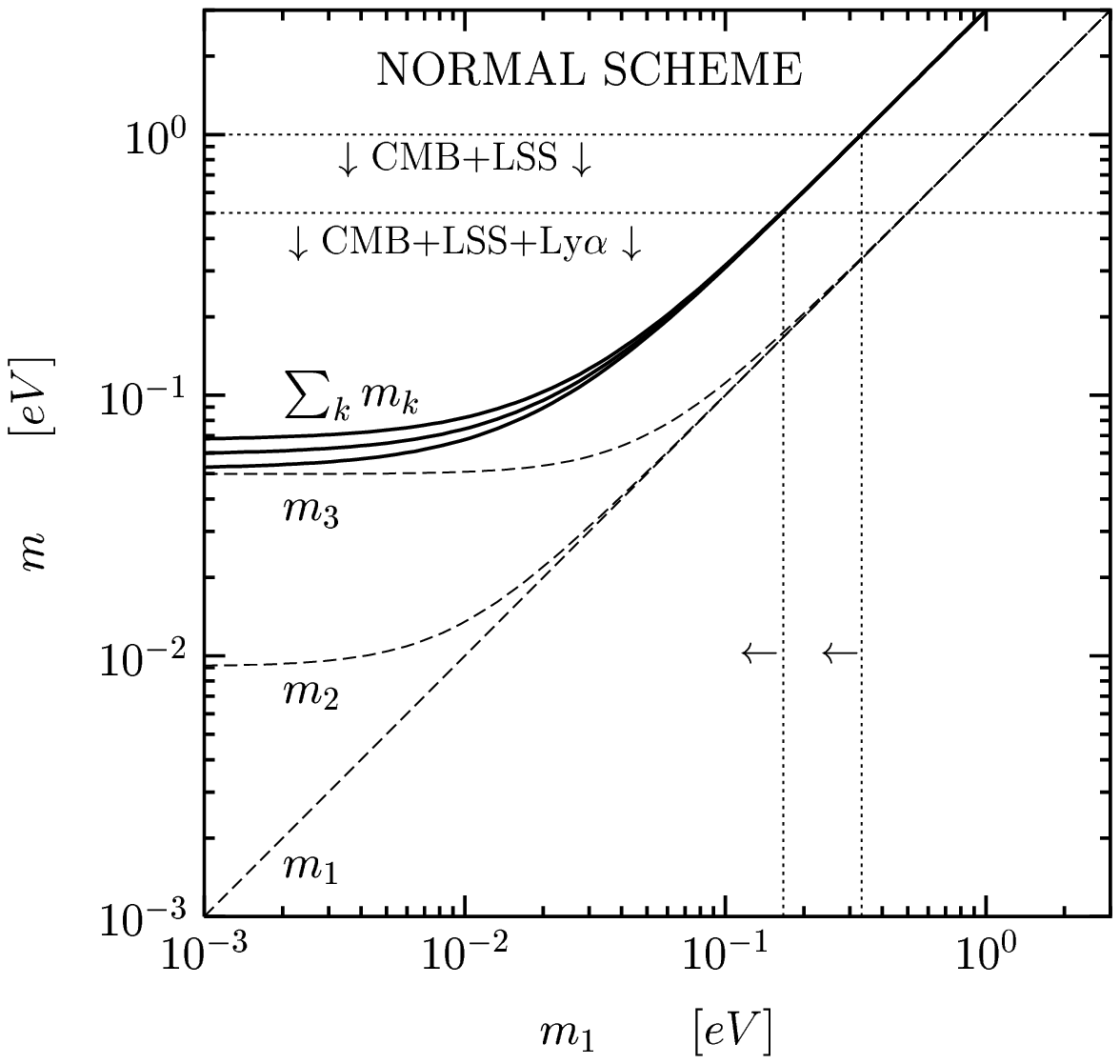}
\end{center}
\end{minipage}
\hfill
\begin{minipage}[t]{0.47\textwidth}
\begin{center}
\includegraphics*[bb=102 430 437 748, width=0.95\textwidth]{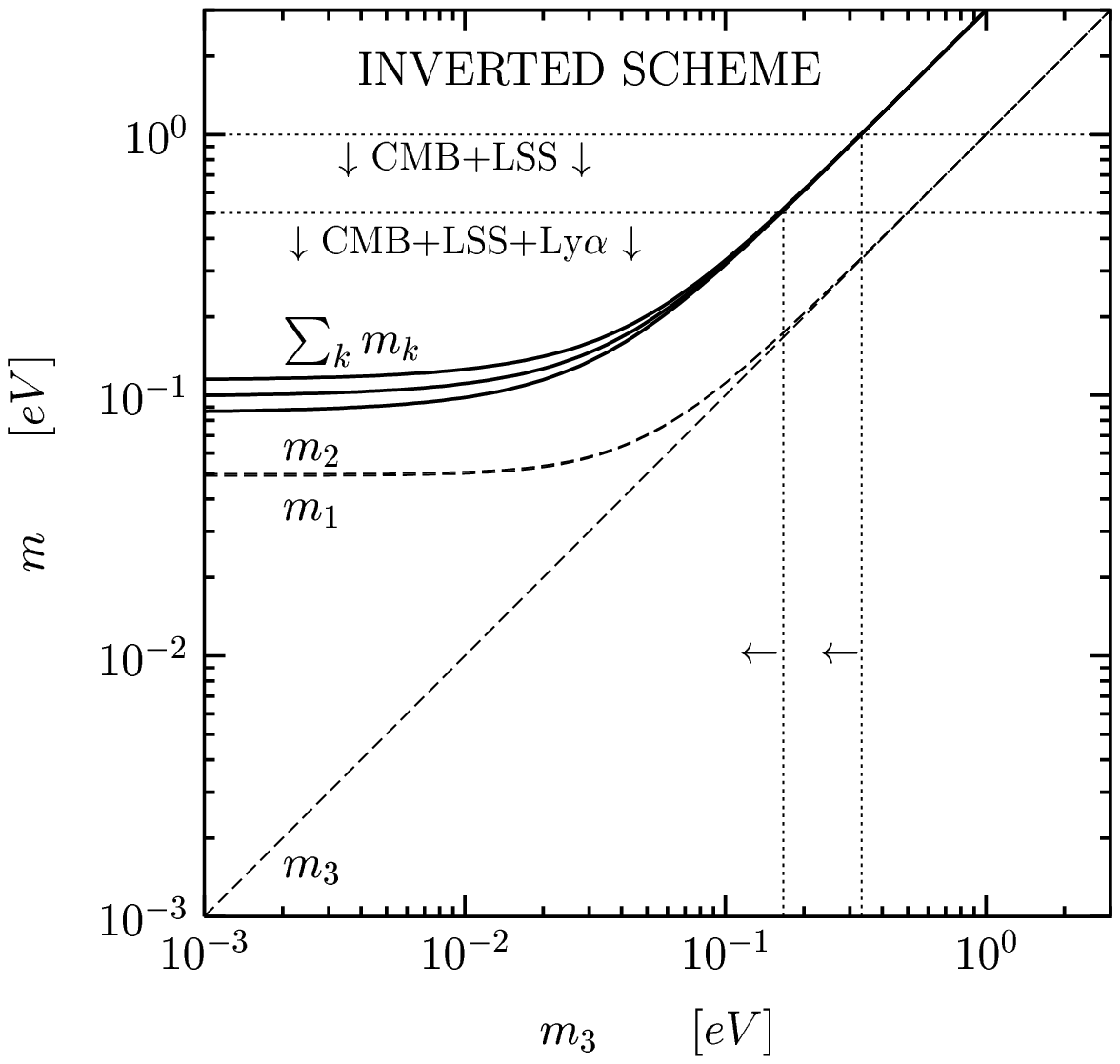}
\end{center}
\end{minipage}
\caption{ \label{cosmo}
Sum of neutrino masses
as a function
of the lightest mass $m_1$ in the normal scheme and $m_3$ in the inverted scheme.
Middle solid lines correspond to the best-fit in Tab.~\ref{fit}.
Extreme solid lines enclose $3\sigma$ ranges.
Dashed lines show the best-fit values of individual masses.
}
\end{figure}

If neutrinos have masses of the order of 1 eV,
they constitute a so-called ``hot dark matter'',
which suppresses the power spectrum of density fluctuations
in the early universe at ``small'' scales, of the order of
1$-$10 Mpc
(see Ref.~\cite{Hu:1998mj}).
The suppression depends on
the sum of neutrino masses
$ \sum_k m_k $.

Recent high precision measurements of
density fluctuations
in the Cosmic Microwave Background
(WMAP)
and
in the Large Scale Structure distribution of galaxies
(2dFGRS, SDSS),
combined with other cosmological data,
have allowed to put stringent upper limits on
$ \sum_k m_k $,
of the order of 1 eV
\cite{Spergel:2003cb,astro-ph/0303076,astro-ph/0303089,astro-ph/0310723,astro-ph/0406594,Seljak:2004xh,hep-ph/0408045}.
However,
different authors have obtained significantly different
upper bounds, mainly because of the different sets of data considered.
The most crucial type of data are the so-called Lyman-$\alpha$ forests,
which are constituted by absorption lines in the spectra of high-redshift quasars
due to intergalactic hydrogen clouds.
Since these clouds have dimensions of the order of
1$-$10 Mpc,
the Lyman-$\alpha$ data are crucial in order to push the
upper bound on $ \sum_k m_k $ below 1 eV.
Unfortunately, the interpretation of
Lyman-$\alpha$ data
may suffer from large systematic uncertainties.
Summarizing the different limits obtained in
Refs.~\cite{Spergel:2003cb,astro-ph/0303076,astro-ph/0303089,astro-ph/0310723,astro-ph/0406594,Seljak:2004xh,hep-ph/0408045},
we estimate the approximate $2\sigma$ upper bounds as
\begin{equation}
\sum_k m_k \lesssim 0.5 \, \mathrm{eV}
\quad
\text{(with Ly$\alpha$)}
\,,
\qquad
\sum_k m_k \lesssim 1 \, \mathrm{eV}
\quad
\text{(without Ly$\alpha$)}
\,.
\label{c01}
\end{equation}
These limits are shown in Fig.~\ref{cosmo},
where we have plotted the value of
$ \sum_k m_k $
as a function
of the unknown value of the lightest mass
($m_1$ in the normal scheme
and
$m_3$ in the inverted scheme),
using the values of the squared-mass differences
in Tab.~\ref{fit}.
One can see that both limits in Eq.~(\ref{c01}) constrain the neutrino masses
in the quasi-degenerate region,
where the upper bound on each individual mass
is one third of the bound on the sum.
In the future, the inverted scheme can be excluded by an upper bound
of about
$8 \times 10^{-2} \, \mathrm{eV}$
on the sum of neutrino masses.

\section{Neutrinoless Double-$\beta$ Decay}
\label{Neutrinoless}

Neutrinoless double-$\beta$ decay is a very important process,
because it is not only sensitive to the absolute value of neutrino masses,
but mainly because it is allowed only if neutrinos
are Majorana particles
\cite{Schechter:1982bd,Takasugi:1984xr}.
A positive result in neutrinoless double-$\beta$ decay
would represent a discovery of a new type of particles,
Majorana particles.
This would be
a fundamental improvement in our understanding of nature.

Neutrinoless double-$\beta$ decays are processes of type
$
\mathcal{N}(A,Z)
\to
\mathcal{N}(A,Z\pm2)
+
e^{\mp}
+
e^{\mp}
$,
in which no neutrino is emitted,
with a change of two units of the total lepton number.
These processes,
which are forbidden in the Standard Model,
have half-lives given by
(see Refs.~\cite{Civitarese:2002tu,hep-ph/0405078})
\begin{equation}
T_{1/2}^{0\nu}
=
\left(
G_{0\nu}
\,
|\mathcal{M}_{0\nu}|^2
\,
|m_{\beta\beta}|^2
\right)^{-1}
\,,
\label{d02}
\end{equation}
where $G_{0\nu}$ is the phase-space factor,
$\mathcal{M}_{0\nu}$
is the nuclear matrix element
and
\begin{equation}
m_{\beta\beta}
=
\sum_k
U_{ek}^2 \, m_k
\label{d03}
\end{equation}
is the effective Majorana mass.

A possible indication of neutrinoless double-$\beta$ decay of ${}^{76}\mathrm{Ge}$
with half-life
\begin{equation}
T_{1/2}^{0\nu}({}^{76}\mathrm{Ge})
=
( 0.69 - 4.18 ) \times 10^{25} \, \mathrm{y}
\qquad
(3\sigma)
\label{d05}
\end{equation}
has been found by the authors of Ref.~\cite{hep-ph/0404088}.
Other experiments did not find any indication of
neutrinoless double-$\beta$ decay.
The most stringent lower bound on $T_{1/2}^{0\nu}({}^{76}\mathrm{Ge})$
has been obtained in the Heidelberg-Moscow experiment \cite{Klapdor-Kleingrothaus:2001yx}:
\begin{equation}
T_{1/2}^{0\nu}({}^{76}\mathrm{Ge})
>
1.9 \times 10^{25} \, \mathrm{y}
\qquad
\text{(90\% CL)}
\,.
\label{d06}
\end{equation}
The IGEX experiment \cite{Aalseth:2002rf}
obtained the comparable limit
$
T_{1/2}^{0\nu}({}^{76}\mathrm{Ge})
>
1.57 \times 10^{25} \, \mathrm{y}
$
(90\% CL).
Hence, the status of the experimental search for neutrinoless double-$\beta$
decays is presently uncertain and new experiments which can check
the indication (\ref{d05}) are needed
(see Ref.~\cite{hep-ph/0405078}).

\begin{figure}[t!]
\begin{minipage}[t]{0.47\textwidth}
\begin{center}
\includegraphics*[bb=119 428 463 751, width=0.95\textwidth]{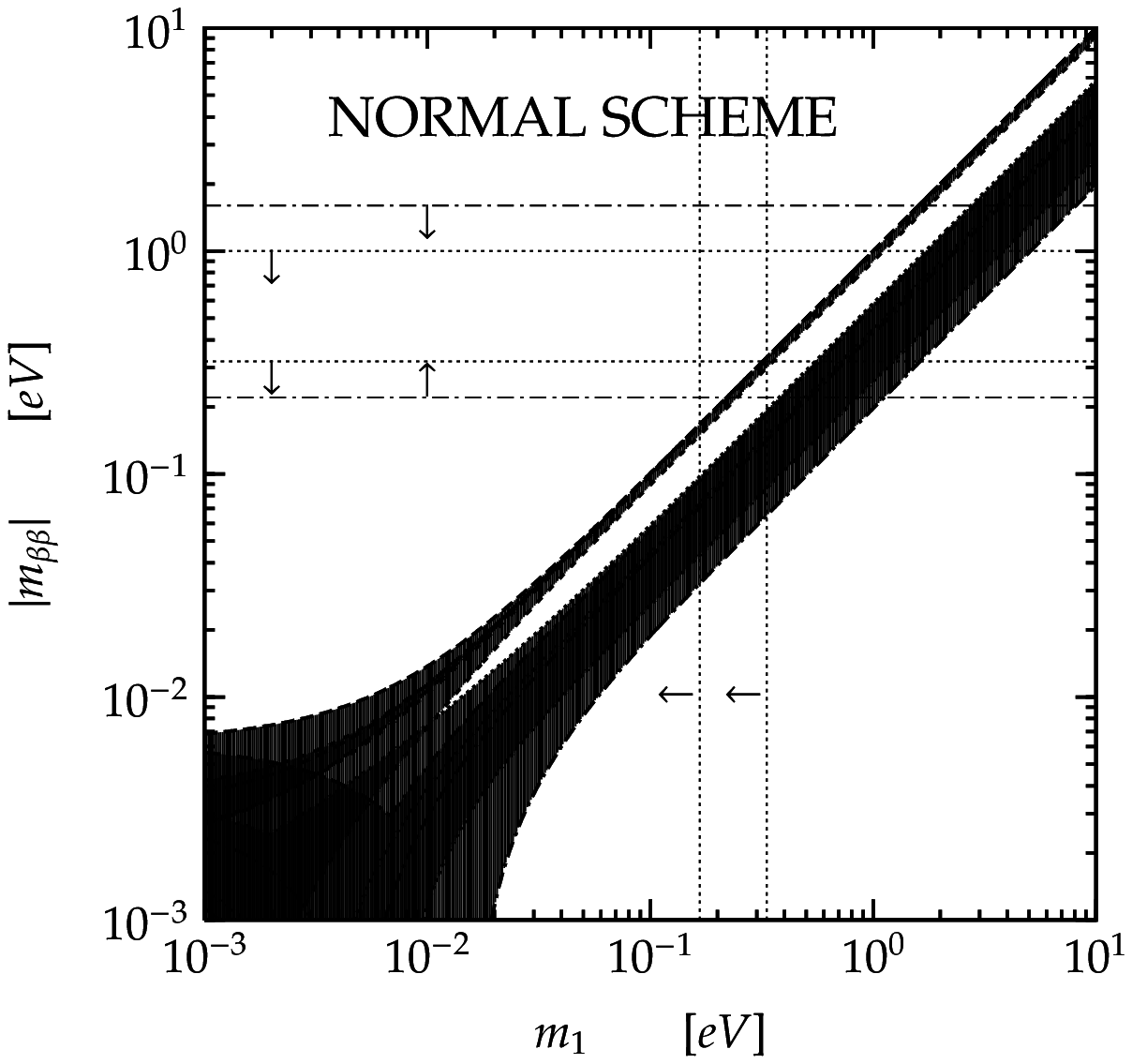}
\end{center}
\end{minipage}
\hfill
\begin{minipage}[t]{0.47\textwidth}
\begin{center}
\includegraphics*[bb=120 427 463 750, width=0.95\textwidth]{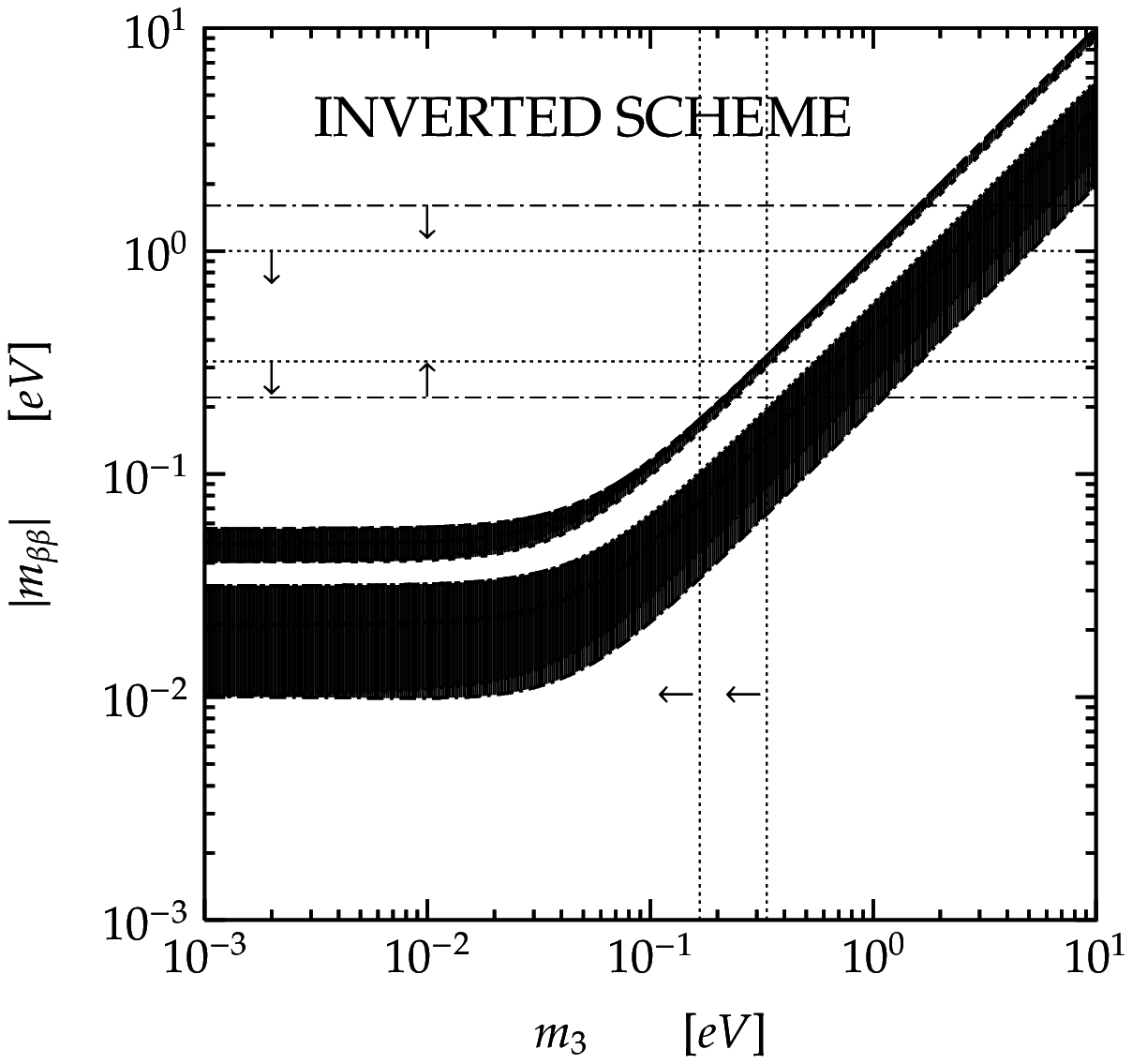}
\end{center}
\end{minipage}
\caption{ \label{db}
Effective Majorana mass $|m_{\beta\beta}|$
in neutrinoless double-$\beta$ decay experiments as a function
of the lightest mass $m_1$ in the normal scheme and $m_3$ in the inverted scheme.
The white areas in the strips need CP violation.
The horizontal dotted lines show the interval (\ref{d09})
of uncertainty of the current experimental upper bound
due to the estimated uncertainty (\ref{d04})
of the value of the nuclear matrix element.
The horizontal dash-dotted lines delimit the range (\ref{d08})
obtained from the indication (\ref{d05}).
The vertical dotted lines correspond to the cosmological upper
bounds on individual neutrino masses in Fig.~\ref{cosmo}.
}
\end{figure}

The extraction of the value of $|m_{\beta\beta}|$
from the data
has unfortunately a large systematic uncertainty,
which is due to the
large theoretical uncertainty in the evaluation of the
nuclear matrix element
$\mathcal{M}_{0\nu}$
(see Refs.~\cite{Civitarese:2002tu,hep-ph/0405078}).
In the following, we will use as a $3\sigma$ range for the
nuclear matrix element
$|\mathcal{M}_{0\nu}|$
the interval which covers the results of reliable calculations
listed in Tab.~2 of Ref.~\cite{hep-ph/0405078}
(other approaches are discussed in Refs.~\cite{Rodin:2003eb,hep-ph/0402250,hep-ph/0403167,hep-ph/0408045}):
\begin{equation}
0.41
\lesssim
|\mathcal{M}_{0\nu}|
\lesssim
1.24
\,,
\label{d04}
\end{equation}
which corresponds to a $3\sigma$ uncertainty of a factor of 3 for the
determination of $|m_{\beta\beta}|$ from $T_{1/2}^{0\nu}({}^{76}\mathrm{Ge})$.
Using the range (\ref{d04}),
the indication (\ref{d05}) implies
\begin{equation}
0.22 \, \mathrm{eV}
\lesssim
|m_{\beta\beta}|
\lesssim
1.6 \, \mathrm{eV}
\,,
\label{d08}
\end{equation}
and the most stringent upper bound (\ref{d06})
implies
\begin{equation}
|m_{\beta\beta}|
\lesssim
0.32 - 1.0 \, \mathrm{eV}
\,.
\label{d09}
\end{equation}

In the standard parameterization of the mixing matrix,
the effective Majorana mass is given by
\begin{equation}
m_{\beta\beta}
=
c_{12}^2 \, c_{13}^2 \, m_1
+
s_{12}^2 \, c_{13}^2 \, e^{i\alpha_{21}} \, m_2
+
s_{13}^2 \, e^{i\alpha_{31}} \, m_3
\,,
\label{003}
\end{equation}
where
$\alpha_{21}$ and $\alpha_{31}$
are Majorana phases
(see Refs.~\cite{hep-ph/9812360,hep-ph/0211462,hep-ph/0310238}),
whose values are unknown.

Figure~\ref{db} shows the allowed range for
$|m_{\beta\beta}|$
as a function
of the unknown value of the lightest mass
($m_1$ in the normal scheme
and
$m_3$ in the inverted scheme),
using the values of the oscillation parameters
in Tab.~\ref{fit}
(see also
Refs.~\cite{Feruglio:2002af-ADD,hep-ph/0304276,hep-ph/0310003,hep-ph/0310238,hep-ph/0402250,Bahcall:2004ip,Petcov-NJP6-109-2004}).
One can see that, in the region where the lightest mass is very small,
the allowed ranges for $|m_{\beta\beta}|$
in the normal and inverted schemes are dramatically different.
This is due to the fact that
in the normal scheme strong cancellations between the contributions
of $m_2$ and $m_3$ are possible,
whereas in the inverted scheme
the contributions of $m_1$ and $m_2$ cannot cancel, because maximal mixing
in the 1$-$2 sector is excluded by solar data
($\vartheta_{12}<\pi/4$ at $5.8\sigma$ \cite{Bahcall:2004ut}).
On the other hand,
there is no difference between the normal and inverted schemes
in the quasi-degenerate region,
which is probed by the present data.
From Fig.~\ref{db} one can see that, in the future,
the normal and inverted schemes may be distinguished by reaching a sensitivity
of about $10^{-2} \, \mathrm{eV}$.

\begin{figure}[t!]
\begin{center}
\includegraphics*[bb=112 430 445 754, width=0.49\textwidth]{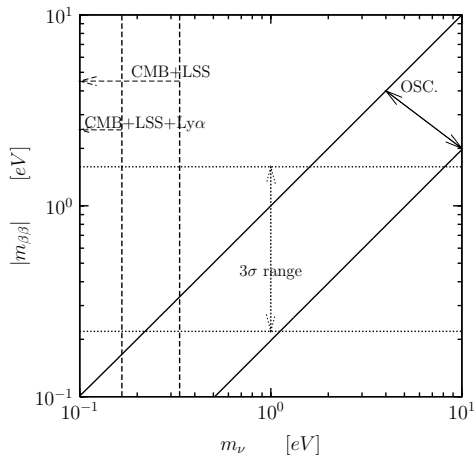}
\end{center}
\caption{ \label{f06}
Allowed regions in the $m_{\nu}$--$|m_{\beta\beta}|$ plane from
the indication of neutrinoless double-$\beta$ decay with the half-life in Eq.~(\ref{d05})
($3\sigma$ range in Eq.~(\ref{d08}) between the dotted horizontal lines),
oscillation data
(allowed range between the diagonal solid lines)
and cosmological measurements
(upper bounds in Eq.~(\ref{c01}) represented by the vertical dashed lines).
The abscissa $ m_{\nu} \simeq m_1 \simeq m_2 \simeq m_3 $ is the scale of quasi-degenerate masses
in both normal and inverted schemes.
}
\end{figure}

In Fig.~\ref{f06} we have enlarged the region of quasi-degenerate masses,
which is practically the same in the normal and inverted schemes,
in order to assess the
compatibility of
the indication of neutrinoless double-$\beta$ decay with the half-life in Eq.~(\ref{d05}),
oscillation data
and cosmological measurements.
One can see that there is a tension
among these sets of data,
especially if the cosmological limit with Lyman-$\alpha$ data is considered
(see also Ref.~\cite{hep-ph/0408045}).

\section{Conclusions}
\label{Conclusions}

The results of neutrino oscillation experiments
have shown that neutrinos are massive particles.
However,
so far we do not know which is the value of the absolute scale of neutrino masses,
except that it is smaller than about 1 eV.
Moreover, we do not know if neutrinos are
Dirac or Majorana
particles.
The solution of these fundamental mysteries
is one of the hot topics in
experimental and theoretical high-energy physics research.

\end{document}